# Autoantibody recognition mechanisms of MUC1

J. C. Phillips

Dept. of Physics and Astronomy, Rutgers University, Piscataway, N. J., 08854

## Abstract

The most cost-effective blood-based, noninvasive molecular cancer biomarkers are based on p53 epitopes and MUC1 tandem repeats. Here we use dimensionally compressed bioinformatic fractal scaling analysis to compare the two distinct and comparable probes, which examine different sections of the autoantibody population, achieving combined sensitivities of order 50%. We discover a promising MUC1 epitope in the SEA region outside the tandem repeats.

**Introduction**  The value of p53 printed 15-mer epitopes as biomarkers for seromic autoantibody paratopes, much more sensitive than complete 393 amino acid (aa) p53, was established in [1], using the ~ 50,000 London patient database. In subsequent papers similar methods identified central MUC1 tandem repeats as also effective, and complementary [2,3]. We have developed thermodynamic scaling methods that utilize only parameter-free bioinformatic scales derived from the geometry of thousands of protein structures recently added to the Protein Data Base [4]. In [5] these scaling methods were tested against the p53 epitopes discovered in [1], with strongly positive results. Here we extend the analysis to MUC1 tandem repeats.

Because p53 is the most studied protein, it is an attractive subject for testing scaling methods, which appear to have universal features, including quantitative aspects of evolution not accessible to phylogenetic trees [4]. Between human and mouse, p53 has evolved with ~ 80% BLAST similarity. Mucin forms multiple families, the ones most relevant to humans being MUC1 and MUC4. Much less is known about mucin than about p53, so we begin by reviewing its salient feature, the 20 aa tandem strict and peripheral repeats near the N terminal of MUC1 and the more variable 16 aa tandem repeats near the N terminal of MUC4. The number of these repeats varies usually from 30 to 90 in human individuals, but is only 16 in mouse [6].

Our scaling methods are focused on quantifying sculpted protein surface topology, making maximal use of modern bioinformatically derived amino acid scales. Modeling with earlier and less accurate scales achieved only limited successes, which have dampened interest in scaling approaches. Our work has encouraged us to suppose that an entire range of new and more accurate results can be achieved quite easily, using computationally trivial EXCEL macros and modern scales. The two scales used here are the hydropathic solvent-accessible surface area scale $\Psi$ based on fractals [4,7], and the beta strand exposed scale $\beta$exp [8,9]. These two scales are strongly anticorrelated (r ~ - 0.9), and their unweighted averages are $<\Psi>$ = 155 and $<\beta$exp$>$ = 145. We plot $\beta$exp* = 300 − $\beta$exp to make the intrinsic surface similarities more obvious.

We are concerned here primarily with detecting interactions of p53 epitopes or mucin fragments with autoantibody paratopes. We previously found good results for p53 epitopes [1,5] using a sliding window wave length W = 9, which is also the lower limit of the fractal $\Psi$ range [7]. By using this value of W, we achieve maximum resolution. In addition, we have examined the correlation r between $\Psi$ and $\beta$exp* for the canonical 1VTSAPDTRPAPGSTAPPAHG20 tandem repeat unit of MUC1, and found a remarkable result. For W = 1, r = 0.72 , but this value increases to 0.90 (0.01) for W = 9 and 11, where it plateaus at the same value as the correlation r of unweighted scales. The effect of using W = 9 for mucin tandem repeats is to recover the strong interscale correlation contained in the original bioinformatic scales. Given the remarkable physiology of mucin, this conclusion is less surprising than it might have been for any other protein. It also strengthens the universality of the bioinformatic results [7,8]. Much lower r values are found for other parts of MUC1, such as the conformationally stressed cleavage domain SEA [9].

We review the p53 results in Figs. 1 and 2, which separate p53 into two parts for clarity. The most effective epitopes are associated with segments that have $\beta$exp* below $\Psi$ by about 10 scale units. Specifically the 40-60 epitope (which could contain a 49-57 9-mer nucleus [5] also is near the deepest hydrophilic minimum (most exposed, and most flexible) sites of p53.

**Results** [2,3] used three MUC1 canonical repeats (60 aa) as a complementary biomarker for p53. They also tried more variable peripheral MUC1 repeats and more approximate MUC4 repeats.

The three MUC1 canonical repeats are less successful than the p53 epitopes, but they complement the p53 epitopes well, and increased the sensitivity from 30% (p53 epitopes alone) to near 50% when combined. The peripheral MUC1 repeats and more approximate MUC4 repeats were less effective. All of these results are similar to the trends seen in MUC1 profiles, shown here only for 2 MUC1 canonical repeats in Fig. 3. Also shown is the effect of an S13A mutation [10]. This should improve binding to autoantibodies, and should be tested.

The most striking feature of Fig. 3 is the very strong correlation r between $\Psi$ and $\beta exp^*$. As mentioned above, this very strong correlation r = 0.90 +/-0.01 occurs with MUC1 repeats only for W $\geq$ 9, and it decreases to 0.82 for W = 5 (a popular choice in older work with less accurate scales [5]) and 0.72 for W = 1. This correlation is unevenly distributed, being closest near the hydrophilic minima. The large number of evolutionarily engineered human MUC1 canonical repeats mutually support the overall strongly hydrophilic region (hydroneutral = 155). They may also enable a 60 aa repeat segment to bind to autoantibody paratopes by enveloping them using weaker hydrogen bonding and van der Waals forces.

The practical question now remaining is whether or not one can find alternatives to MUC1 repeats that could be used in place of the MUC1 (n = 3) segments used in [2,3]. If one assumes that the main feature for binding to autoantibodies is a strong $\beta$ strand interaction, and that this is improved by proximity to a flexible hydrophilic $\Psi$ minimum, then these two conditions are sufficient to eliminate peripheral MUC1 repeats and more approximate MUC4 repeats. One can still look at other species. Bovine MUC1 repeats turn out to have weak $\beta$ strand interactions, so only mouse remains. They deserve consideration because murine tumor models are widely used as tests for various applications, including predicting tumor mutations [11].

Most of the mouse repeats have only weak $\beta$ strand interactions, but near the end of the 16 repeats, there is a strong interaction in repeat 15 (Uniprot Q02496), sites 327-346. When this repeat is supported by repeats 14 and 16, the profile shown in Fig. 4 is obtained. There is a 7 aa region PDHNGSS (332-338) that has strong $\beta$ strand interactions, but its average $\beta exp^*9$ value below $\Psi MZ9$ is only 4. Overall mouse MUC1 repeats have only weaker $\beta$ strand interactions than MUC1 canonical repeats.

The gap between ΨMZ9 and βexp*9 in the 332-338 valley can be increased to 7 by three S>A mutations (Fig. 5). The resulting well is more favorable for β strand interactions, but unlike the 40-55 and 45-60 p53 epitopes in Fig. 1, it is not adjacent to a flexible deep hydrophilic Ψ minimum. This absence of flexibility could reduce assay sensitivity.

Another possibility is to go outside the repeat domain into the SEA domain 1038-1148, which is cleaved at 1097. Its profiles are shown in Fig. 6, because the 9-mer 24-32 (1062-1070, STDYYQELQ) has strong β strand interactions. The average 24-32 difference between ΨMZ9 and βexp*9 is 16.8 (larger than MUC1 canonical repeats and p53 epitopes). Expanding the length to 11-mer 23-33 reduces the difference to 14.4 (still large). However, this sequence itself is disordered (like the rest of MUC1), and it may not be topologically stable [12].

An NMR study in solution found an α helix (1063-1081) and β strands (1042-1049, 1086 – 1096) in an SEA domain including an N-terminal $His_6$ tag [9]. This suggests printing the truncated $His_6$1041-1077, whose profiles are shown in Fig. 6, with a mutation S1062A [12]. Over the range 1062-1070) the mutation increases the average 24-32 difference between ΨMZ9 and βexp*9 to 18.5. For the more stable conditions of an ELISA assay [1-3], it might be possible to move the $His_6$ tag closer to the flexible hydrophilic minimum near 1025, enhancing the β strand propensities further.

**Conclusions** We found a striking increased correlation in disordered mucin canonical repeats between hydropathic fractal shaping Ψ [7] and β strand propensities [8] in the fractal length range W ≥ 9. It appears that this unique correlation is the molecular key to the mucin tandem repeat functionality, which may include autoantibody recruitment. All the conclusions of [1-3] are consistent with the β strand interaction model, previously proposed for p53 epitopes [5]. O-glycosylated mucin fragments are less sensitive as biomarkers than bare MUC1 core repeats [2,3], probably because glycosylation weakens β strand hydrogen bond interactions.

Serine O-glycosylation increases the van der Waals packing volume of the Ser site, and S13 O-glycosylation is the first step towards colon cancer (Fig. 1d,e of [2]). Conversely an S13A mutation decreases that volume, so linear extrapolation suggests this would increase biomarker sensitivity. However, human proteins have generally evolved to be very close to functional criticality [4], and the large number of human tandem repeats (compared to mouse or bovine) suggests that mucin autoantibody recruitment is near criticality. In that case, any change in the canonical repeat, either by Ser O-glycosylation or Ser alanining, will reduce sensitivity.

A 9-mer has been identified in the MUC1 SEA domain as having strong $\beta$ strand interactions. It is close to a hydrophilic hinge, and is a potential biomarker. The 9-mer has choppy $\Psi MZ9$ and $\beta exp*9$ profiles, so this possibility is remote, but still worth testing.

**Methods** The MUC1 repeats and SEA sequences are identified in Uniprot P15941. The canonical MUC1 sequence PDTRPAPGSTAPPAHGVTSA varies in the peripheral tandem regions [12]. The $\beta exp*$ and $\Psi MZ$ scales used here are listed in [13,14]. The calculations described here are very simple, and are most easily done on an EXCEL macro. The one used in this paper was built by Niels Voorhoeve and refined by Douglass C. Allan. Many of the author's published papers referenced here are conveniently grouped together at arXiv, Quantitative Biology, J. C. Phillips.

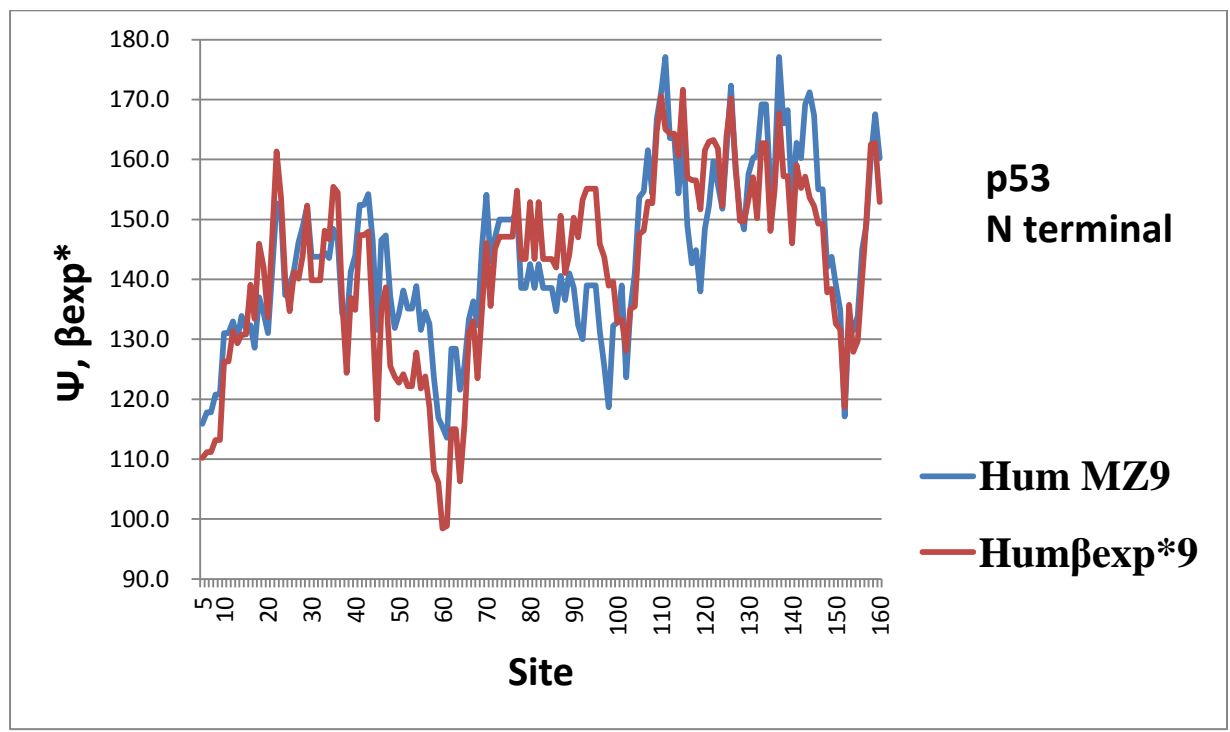

Fig. 1. The strongest signal for early cancer detection in [1] came from the 15 mers corresponding to 40-55 and 45-60. It was suggested in [5] that both signals could have been caused by a common 9 or 11 mer near 45-55 = LSPDDIEQWFT, with $<\Psi MZ - \beta exp*> \sim 10$. This region is strongly hydrophilic ($<\Psi MZ> \sim 137$). Bearing in mind the difference of 10 between $<\Psi MZ>$ and $<\beta exp*>$, it is equally strong in its preference for exposed beta strands. It is possible that this epitope occurs as an early cancer signal, but does not occur with tumors, because it lies in the N terminal region, away from the DNA binding region.

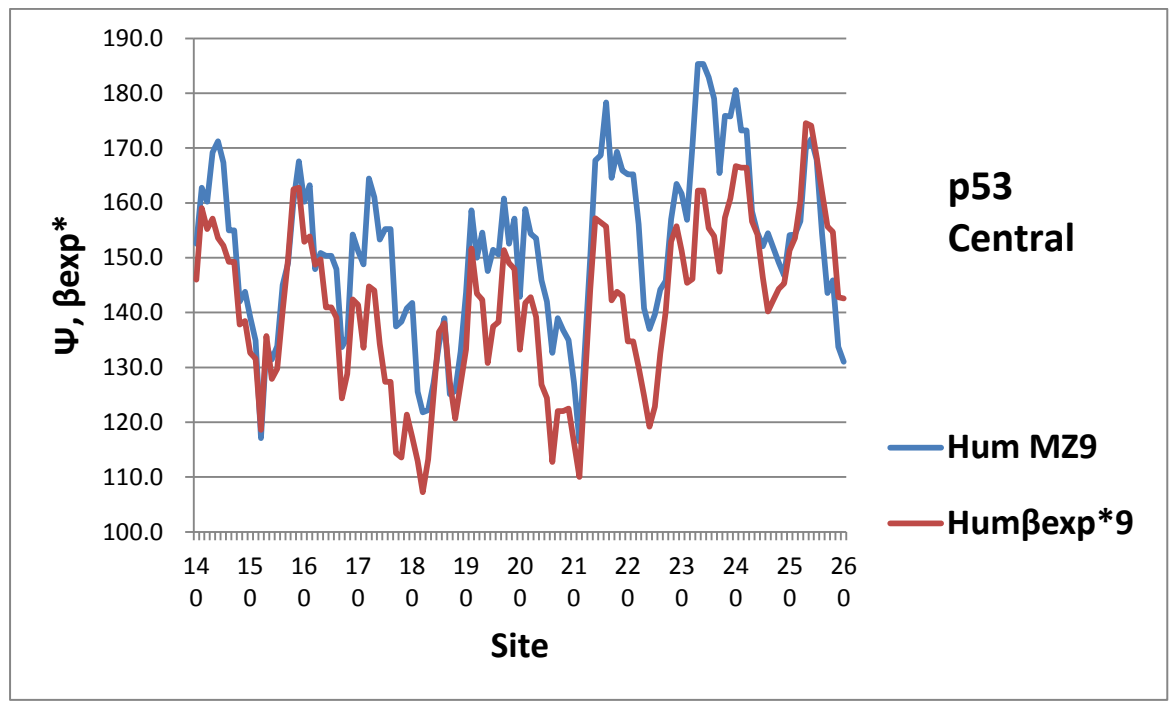

Fig, 2. Here is shown the central part of the DNA binding region 102-292. The strong 15-mer epitope for colon tumors is 165-180. This is the region that is most similar to 40-60 in Fig. 1, because it is both strongly hydrophilic ($\Psi$MZ), equally strong in its preference for exposed beta strands ($\Psi$MZ - $\beta$exp* ~ 10). It is possible that this epitope occurs in tumors but not as an early cancer signal because it lies in the DNA binding region.

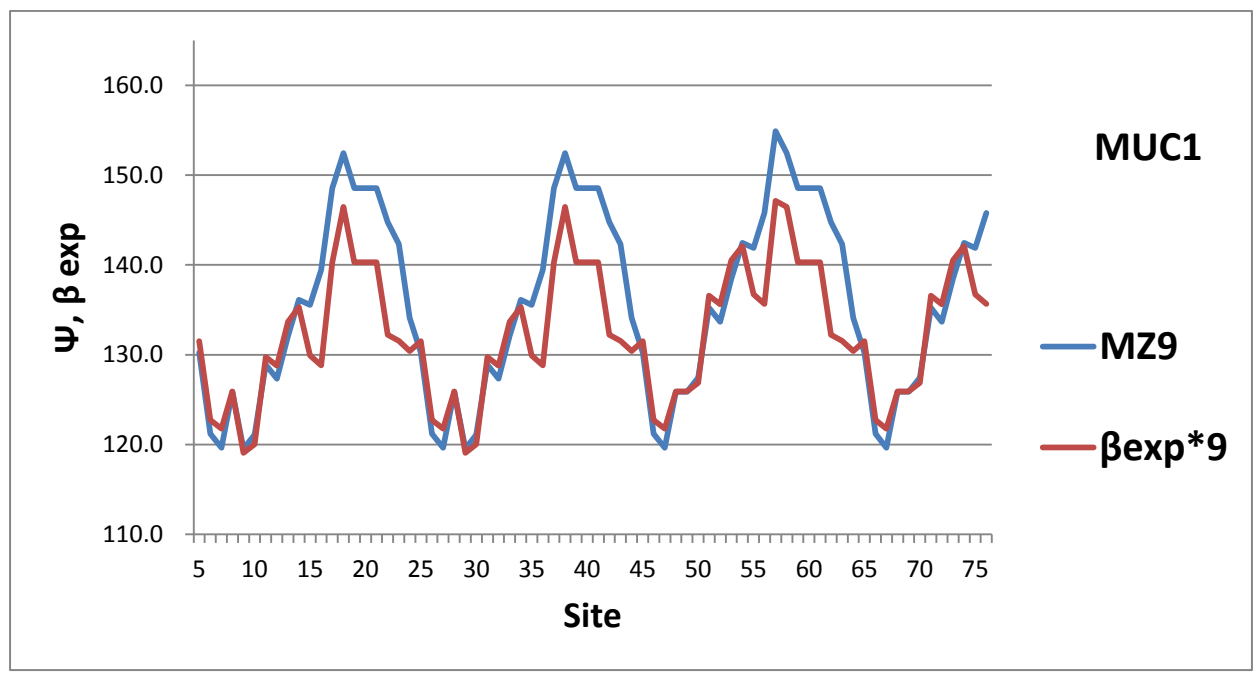

Fig. 3. For the two canonical (or "core") MUC1 repeats (1-40) the βexp* values lie below the ΨMZ values (W = 9), but in a special way: about 10 near the relatively hydrophobic maxima, but near 0 for the extremely hydrophilic minima. The two mutated core repeats S13A (41-80) increase the overall gap, by making the ΨMZ profile more hydrophobic, which is expected to increase binding to autoantibodies. The S13A mutation also makes the hydrophilic minimum narrower, and reduces flexibility. Overall the MUC1 ΨMZ values are hydrophilic, and lie below the hydroneutral value for ΨMZ of 155. This is consistent with the overall disordered mucin structures, and is similar to the overlapping 15-mer 40-60 region of Fig. 1.

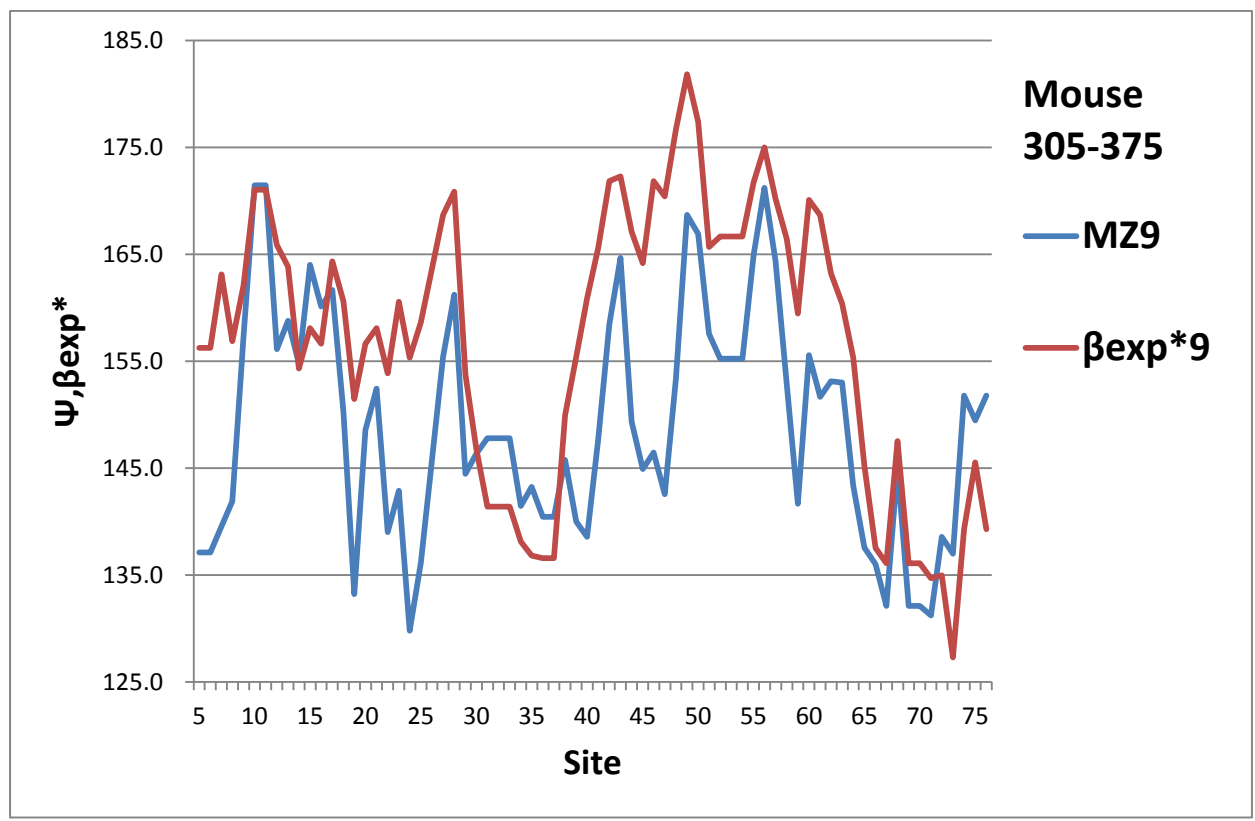

Fig. 4. The β strand interactions are weak across most mouse repeats, but near the C terminal end of the repeat region, a narrow favorable range is found, extending from 331 to 337.

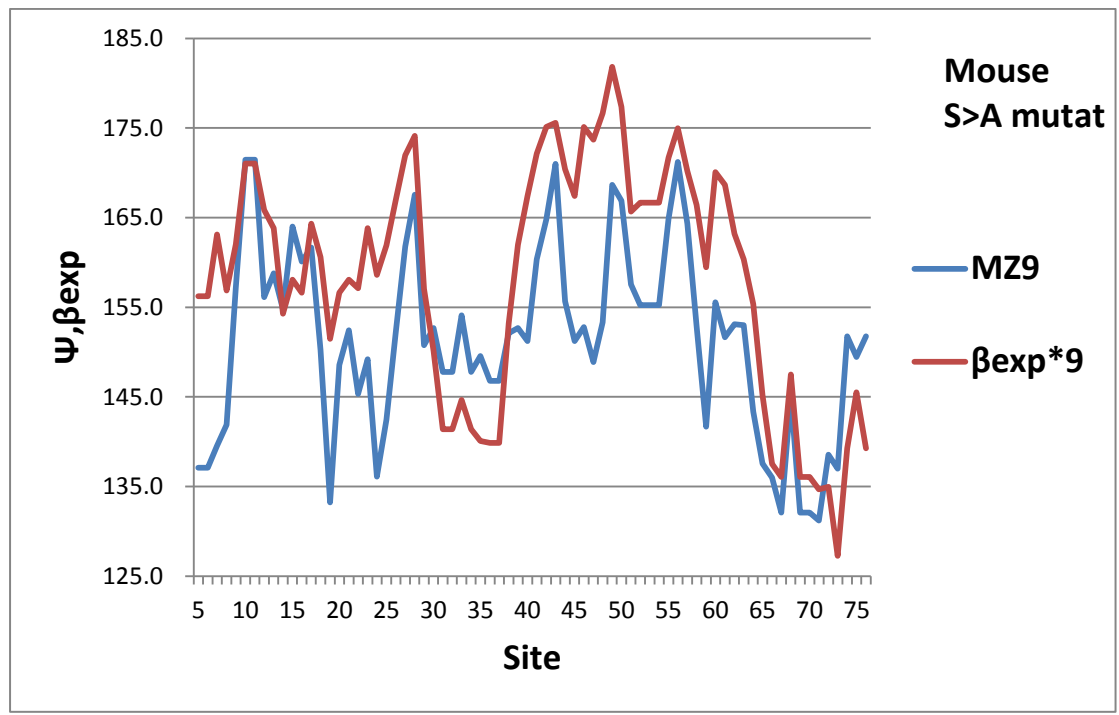

Fig. 5. The difference between ΨMZ9 and βexp*9 in the 331-337 well is increased by S326A, S337A and S343A mutations.

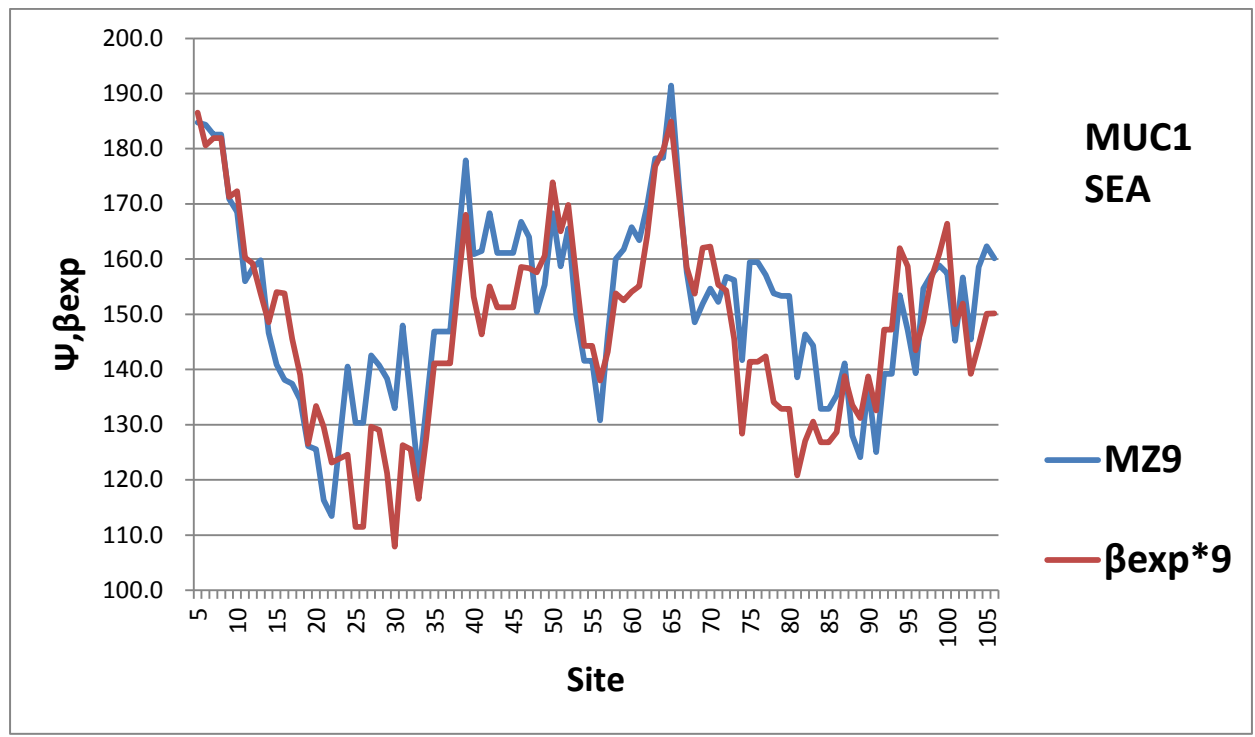

Fig. 6. The SEA MUC1 human domain cleaves proteolytically at site 61 (Uniprot 1061-1969), near the central hydrophilic minimum. The deepest hydrophilic minimum is near site 22. The 9-mer 24-32 has strong β strand interactions, but these oscillate, suggesting a possible topological instability.

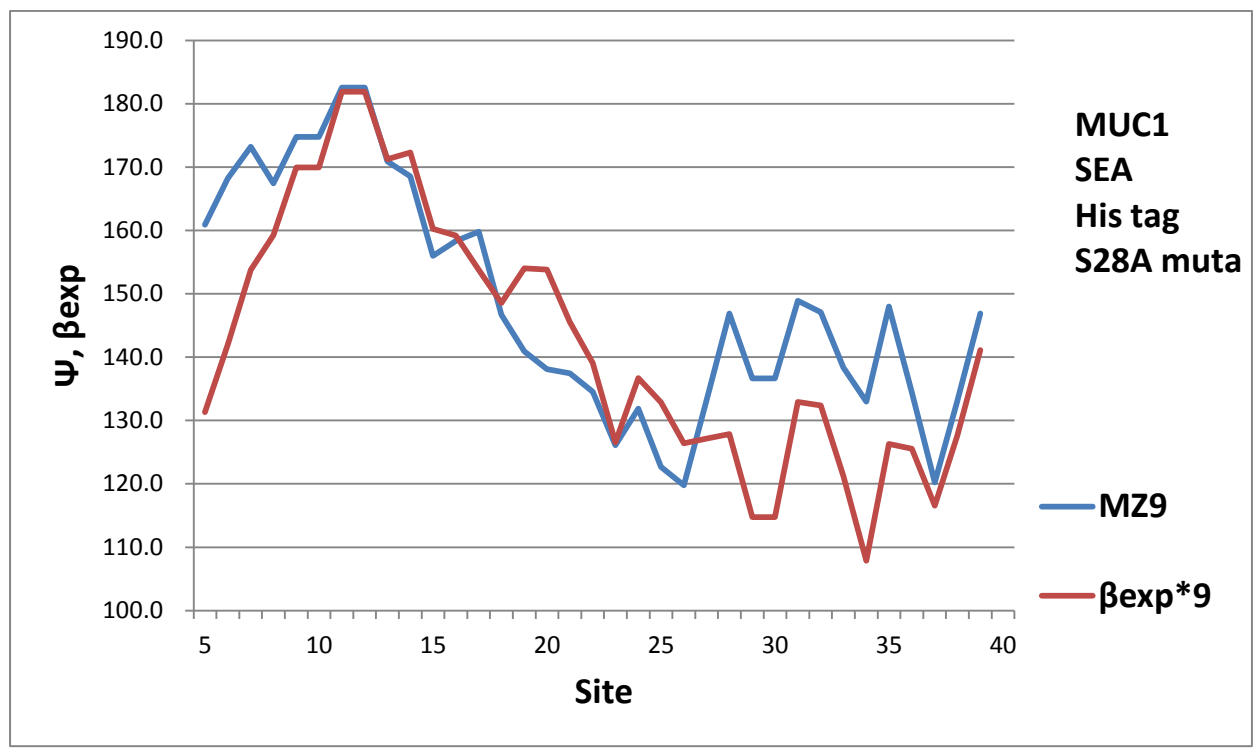

Fig. 7. The truncated SEA domain with an N terminal His6 tag. Sites 24-26 correspond to Uniprot P15941 LED 1058-1060. Strong β strand propensities occur across the 28-36 range, corresponding to 32 in Fig. 6. This profile includes an additional S1062A mutation.